\documentclass[11pt,a4paper]{article} 

\usepackage{jcappub}
\usepackage{graphicx}

\title{Observational consequences of the Standard Model Higgs inflation variants}

\author{Lucia Popa} 

\affiliation{Institute for Space Sciences,\\ 
Bucharest-Magurele, Ro-077125 Romania} 

\emailAdd{lpopa@spacescience.ro} 

\abstract{ We consider the possibility to observationally differentiate  
the Standard Model (SM) Higgs driven inflation with non-minimal coupling 
to gravity from other variants of SM Higgs inflation 
based on the scalar field theories with non-canonical kinetic term 
such as Galileon-like kinetic term and kinetic term with non-minimal derivative coupling to the Einstein tensor.\\
In order to ensure consistent results, we study the SM Higgs inflation variants 
by using the same method, computing the full dynamics of the background and perturbations of the Higgs field during inflation at quantum level.\\
Assuming that all the SM Higgs inflation variants are consistent theories, 
we use the MCMC technique to derive constraints on the inflationnoary parameters and the Higgs boson mass from their fit to WMAP7+SN+BAO data set.  
We conclude that a combination of a Higgs mass measurement by the LHC and accurate determination by the PLANCK satellite of the spectral index of curvature perturbations  
and tensor-to-scalar ratio will enable to distinguish among these models. 
We also show that the consistency relations of the 
SM Higgs inflation variants are distinct enough to differentiate the models.}


\def\be{\begin{equation}}
\def\ee{\end{equation}}
\def\ba{\begin{eqnarray}}
\def\ea{\end{eqnarray}}
\usepackage{amssymb,graphicx,subfigure}
\begin{document}
\begin{flushright}
\end{flushright}

\maketitle

\section{Introduction}

There has been much interest recently in models of inflation 
in which the Standard Model  (SM) Higgs boson 
non-minimally coupled to the  Ricci scalar  
can give rise to inflation 
without the need for additional degrees of freedom 
to the SM 
\cite{BezSha08,BarSta08,BerSha09,deSimone09,BGS09}. 
This scenario is based on the observation that the problem of 
the very small value of the Higgs quadratic coupling required  by the
Cosmic Microwave Background (CMB) anisotropy data can be solved if the Higgs 
inflaton has a large non-minimally coupling with gravity \cite{BarKa94}. 
The resultant Higgs inflaton  effective  potential 
in the inflationary domain is effectively flat and can result in successful 
inflation for values of non-minimally coupling constant  $\zeta \sim 10^4$  ($\zeta$-inflation), 
allowing for cosmological values for Higgs boson mass in a window in which the electroweak vacuum  
is stable and therefore sensitive to  the field fluctuations during the early stages of the universe \cite{Espinosa08}. \\
Limits of validity of $\zeta$-inflation have been recently debated by several authors. Specifically, Barbon \& Espinosa (2009) argued that the large coupling of Higgs inflaton to the Ricci scalar 
makes this model invalid beyond the ultraviolet cutoff scale 
$\Lambda_{\zeta} \simeq M_{P}/\zeta$ (hereafter $M_P=2.4 \times 10^{18}$ GeV is the reduced Planck mass) which is below the Higgs field expectation value  at $\cal N$ {\it e}-foldings during inflation, $\phi \simeq \sqrt{N}M_P/\sqrt{\zeta}$. As consequence,
at the ultraviolet cutoff scale  $\Lambda_{\zeta}$ 
at least one of the cross-sections of different scattering processes hits the unitarity bound \cite{Burgess09}.  
The fact that the quantum corrections due to the strong 
coupling to gravity makes the perturbative analysis to break down at energy scales above $\Lambda_{\zeta}$ was interpreted as a signature of a new physics, implying higher dimensional operators at energies above $\Lambda_{\zeta}$. 
Few authors addressed the issue of  $\zeta$-inflation
naturalness with respect to perturbativity and unitarity violation in 
Jordan and Einstein frames \cite{Lerner10a,Lerner10b,Burgess10,Hertzberg10}. They show that  
the apparent breakdown of the theory in the Jordan frame does not imply new physics, 
but a failure of the perturbation theory in the Jordan frame as a calculation method.
These works demonstrate that, for inflation based on a single scalar field 
with large non-minimal coupling, the quantum corrections at high energy scales 
are small, making the perturbative analysis valid and  consequently there is no a breakdown of unitarity at the energy scale $\Lambda_{\zeta}$. \\
Recent works  \cite{Bez11,Ferrara11,Atkins_a}
revisit the  issue  of self-consistency of  $\zeta$-inflation model emphasizing that the scale of unitarity violation depends on the size of the Higgs background field. 
They found the cutoff scale $\Lambda_{\zeta}$  higher than the relevant dynamical scales 
throughout the whole history of the universe, including the inﬂationary epoch and reheating, indicating that the theory does not violate unitarity at any time.\\
Although the above conclusion theoretically may allow to work within the regime of validity of the effective theory for inflationary calculations, other variants of Higgs inflation within the SM have been proposed. 
These models are based on scalar field theories with non-canonical kinetic term \cite{Picon} and are motivated by some particle physics models of inflation such us the Dirac-Born-Infed inflation \cite{Ali,Easson}. 

A model with non-minimal derivative
couplings was proposed in \cite{Amendola,Capo_a,Capo_b} in context of inﬂationary cosmology, and recently, the inflation driven by a scalar field with a Higgs potential
and non-minimal kinetic couplings to itself and to the curvature 
was proposed in \cite{Granda_a,Granda_b}. 
In the new model of Higgs inflation (E-inflation)
an enhanced kinetic term in the Lagrangian
propagating no more degrees of freedom than the minimally coupled scalar
field in General Relativity (GR) is obtained by considering non-minimal
derivative coupling of the SM Higgs boson to the Einstein tensor \cite{Germani10a}.
The power counting analysis indicates that the scale at which the unitarity bound is violated in this model is $\Lambda (H) \simeq (2H^2 \kappa)^{1/3}$, where $H$ is
the Hubble expansion during inflation. Requiring the unitarity constraint $H \ll \Lambda$ to be satisfied, it is shown that the non-minimal derivative coupling
allows a Higgs boson self-coupling value $\lambda$ within the limits expected from the
collider experiments while the cosmological perturbations in this model are  consistent
with present cosmological data \cite{Germani10b,Atkins_b}. 

The Galileon models of inflation (G-inflation) 
are constructed by introducing a scalar field with a self-interaction whose Lagrangian 
is invariant under Galileon symmetry  $\partial_{\mu}\phi \rightarrow \partial_{\mu}\phi +b_{\mu}$, which maintains the equation of motion of the scalar field as a second-order differential equation, preventing the theory from exhibiting new degrees of freedom and avoiding ghost or instability problems  \cite{Nicolis09,Deffayet09a,Deffayet09b}.
Recently, an inflation model in which inflation is driven by the SM Higgs scalar field with a Galileon-like kinetic term  has been proposed \cite{Koba10}. 
The dynamics of background and perturbations of G-inflaton show new features 
brought by the modified kinetic term in the Lagrangian, when compared with the standard slow-roll inflation, such as: violation of the standard consistency relation \cite{Koba10,Kama10}, violation of the Null Energy Condition \cite{Creminelli10},
generation of isocurvature perturbations and large primordial non-Gaussianity \cite{Mizuno10,Burrage11,Felice11a,Koba11}. 

In this paper we analyze to possibility to distinguish among 
the SM Higgs inflation variants by using the existing cosmological and astrophysical 
measurements. We will adopt a similar philosophy as in  \cite{Lerner11},
considering that the SM Higgs inflation variants are consistent theories  and  therefore should be subject to rigorous tests against experimental data.\\ 
Cosmological constraints on the SM Higgs driven  $\zeta$-inflation
have been discussed by a number of authors \cite{BarSta08,BerSha09,deSimone09}.  
These constraints are based on mapping between the Renormalization Group (RG) 
flow equations and the spectral index  of the curvature perturbations obtained by the WMAP team, parametrized in terms of the number of ${\it e}$-foldings till the end of inflation. In a recent paper \cite{Popa10} we studied the implications  of the  
full dynamics of background and perturbations of the Higgs field during 
 $\zeta$-inflation at quantum level and  derive constaints on the inflationary parameters and the Higgs boson mass from the analysis  of WMAP 7-year
CMB measurements complemented  with astrophysical distance measurements 
\cite{Komatsu10,Larson}. \\ 
Our goal in this paper is to 
refine the computation for $\zeta$-inflation 
by considering in the analysis the improved RG flow equations, accounting at the same time 
for the possible uncertainties in the theoretical determination of the 
reheating temperature. We also make a similar analysis for  G-inflation and E-inflation models and derive constraints on the inflationary parameters and Higgs boson mass from their fit to the same data set, ensuring in this way that the differences in the predictions of the observables are due to the differences in the underlying theories of the SM Higgs inflation and therefore can be used to distinguish among them.  

The paper is organized as follows. 
In Section~2 we introduce the there models of the SM Higgs inflation that we consider: 
$\zeta$-inflation \cite{BezSha08}, G-inflation \cite{Koba10} and E-inflation \cite{Germani10a,Germani10b}. In Section~3 we compute the  RG improved  Higgs  field potential
and in Section~4 we present our main results. In Section~5 we draw our conclusions. \\
Throughout the paper we consider an homogeneous and isotropic flat background described by the Friedmann-Robertson-Walker (FRW) metric: 
\begin{equation}
\label{FRW}
{\rm d}s^2=g_{\mu,\nu}\,{\rm d}x_{\mu}\,{\rm d}x^{\nu}  =  -{\rm d}t^2+a^2(t){\rm d}x^2 \,, 
\end{equation}
where $a$ is the cosmological scale factor, 
$\kappa^2 \equiv 8 \pi M_{pl}^{-2}$ (where $M_{pl}\simeq 1.22 \times 10^{19}$ GeV is the Planck mass), 
overdots denotes the time derivatives and $_{,X} \equiv \partial/ \partial {X} $. \\
We also remind that
the canonical tree-level action describing the SM Higgs inflation is given by:
\begin{equation} 
\label{action0} 
S_0=\int{{\rm d}^4\,x}\sqrt{-g}\, \left[\frac{1}{2\kappa^2}{\bf R}+
{\cal L}_{SM}\right]\,,
\end{equation}
where ${\bf R}$ is the Ricci scalar and ${\cal L}_{SM}$ is the tree-level SM Higgs Lagrangian:
\begin{equation}
\label{L_SMG}
{\cal L}_{SM}= g^{\mu \nu}{\cal D}_{\mu}{\cal H}^{\dagger}{\cal D}_{\nu}{\cal H} 
-\lambda \left({\cal H}^{\dagger}{\cal H} -v^2\right)^2 \,.
\end{equation}
In the above expression ${\cal H}$ is the Higgs boson doublet, ${\cal D}_{\mu}$ is the 
covariant derivative with respect to $SU(2) \times SU(1)$ and $v$ is the vacuum expectation value 
(vev) of the Higgs broken phase of the SM.  Rotating the Higgs doublet such that 
${\cal H}^T=(1/\sqrt{2})(0,v+\phi)$ 
and assuming that $\phi$ is much greather than vev during inflation we have:
\begin{eqnarray}
\label{L_SM}
{\cal L}_{SM}=P(\phi,X)=X-V(\phi)\,, 
\hspace{1cm} 
\label{pot}
X=-\frac{1}{2}g^{\mu\nu}\partial_{\mu}\partial_{\nu}\phi \,,
\hspace{1cm} V(\phi)=\frac{\lambda}{4}\phi^4\,,
\end{eqnarray}
where $\lambda$ is the tree-level Higgs self coupling and $V(\phi)$ is the corresponding Higgs potential. 

\section{Models: background equations and cosmological perturbations}

In this section we derive the background equations of motion of the scalar field 
and present the power spectra of scalar and tensor perturbations derived in the framework of linear cosmological perturbation theory for the SM Higgs inflation models that we consider.
A review of chaotic inflation formalism in modified gravitational theories can be found in \cite{Felice11b}.

\subsection{Higgs driven $\zeta$-inflation } 

The general action for these models in the Jordan frame is given by 
\cite{Futamase89,Felice11b}:
\begin{equation}
\label{action}
S_{J} \equiv \int d^4 x \sqrt{-g} \left[  U(\phi) {\bf R}
            -\frac{1}{2}G(\phi)(\nabla \phi)^2 -V(\phi)\right] \,,
\end{equation}
where $U(\phi)$ is a general coefficient of the Ricci scalar, {\it \bf R}, giving rise to the non-minimal coupling, $G(\phi)$ is the general coefficient of kinetic energy and $V(\phi)$ 
is the general potential. \\
The generalized $U(\phi)${\it \bf R} gravity theory in Equation (\ref{action}) 
includes diverse cases of coupling. For generally coupled scalar field 
$U=(\gamma + \kappa^2\zeta \phi^2)$,  $G(\phi)=1$ and 
$\gamma$ and $\zeta$ are constants. The non-minimally coupled scalar field is the case 
with $\gamma=1$ while the conformal coupled scalar field is the case with 
$\gamma=1$ and $\zeta=1/6$. 
The conformal transformation for the action given  in Equation (\ref{action}) can be achieved by defining the Einstein frame line element:
\begin{eqnarray}
\label{RW}
{\rm d}{\hat s}^2  
  =  \Omega \left(-{\rm d}{\hat t}^2+\hat{a}^2({\hat t}){\rm d}x^2\right)\,,
\hspace{0.5cm}
\Omega  = 2 \kappa^2 U(\phi)\,, 
\hspace{0.5cm} U(\phi)=(1+\kappa^2\zeta \phi^2) \,,
\end{eqnarray}
where the quantities in the Einstein frame are marked by caret.
From the above equations 
we get:
\begin{eqnarray} 
\label{ta}
{\rm d}\hat{a}=\sqrt{\Omega}\,{\rm d}a\,, \hspace{1cm} 
{\rm d}\hat{t}= \sqrt{\Omega}\, {\rm d}t \,,
\end{eqnarray}
and scalar potential $\hat{V}(\hat{\phi})$ in the Einstein frame is  given by: 
\begin{equation}
\label{Ve}
\hat{V}(\hat{\phi})  =  \frac{1}{4 \kappa^4}\frac{V(\phi)}{U^2(\phi)}\,.
\end{equation}
The non-minimal coupling to the gravitational field introduces a modification of the Higgs field propagator by a suppresion factor $s(\hat{\phi})$ defined as \cite{deSimone09,BarSta09}:
\begin{equation}
\label{dphi}
s(\hat{\phi})^{-2}=\left( \frac{{\rm d}\hat{\phi}}{{\rm d}\phi}\right)^2  =  
\frac{1}{2 \kappa^2}\frac{G(\phi) U(\phi)+3 U^2(\phi)_{,\phi}}
{  U^2(\phi)} \,,
\end{equation}
that made the kinetic energy in the Einstein frame canonical with respect to the  the new scalar field ${\hat \phi}$. \\
{\it Background equations.}
The Friedmann equation in the Einstein frame reads as \cite{Komatsu98,Tsu04}: 
\begin{equation}
\label{ein_eq}
\hat{H}^2  =  \frac{\kappa^2}{3}\left[\left(\frac{{\rm d}\hat{\phi}}{{\rm d}\hat{t}}\right)^2 + 
\hat{V}(\hat{\phi})\right]\,, \\
\end{equation} 
where:
\begin{eqnarray}
\label{hub}
\hat{H}  \equiv  \frac{1}{\hat{a}} 
\frac{ {\rm d}\hat{a}} {{\rm d}\hat{t}} &  = &  \frac{1}{\sqrt{\Omega}}
\left[ H+ \frac{1} {2 \Omega} \frac{ {\rm d} \Omega} { {\rm d} t } \right]\,,\\
\label{field}
\frac{ {\rm d}\hat {\phi}} {{\rm d} \hat{t}}  & = &
\left( \frac{ {\rm d} \hat{\phi}} { {\rm d} \phi } \right)
\left( \frac{ {\rm d} t } { {\rm d} \hat{t} } \right) \dot{\phi}    
\end{eqnarray}
Equations (\ref{hub}) and (\ref{field}) are enough to compute the background 
field evolution in the Einstein frame once the field equations in the Jordan 
frame are known. \\
The Jordan-frame field equations from action (\ref{action}) are obtained as \cite{Komatsu99}:
\begin{equation}
H^2  =  \frac{\kappa^2}{3\,(1+\kappa^2\zeta\phi^2)} 
\left[V(\phi) + \frac{1}{2} \dot{\phi}^2- 6\,\zeta H \phi \dot{\phi} \right]\,,   
\end{equation}
\begin{eqnarray}
\ddot{\phi}  + 3 H \dot{\phi}  +  \left( \frac{ \kappa^2 \zeta \phi^2 (1+6 \zeta) } {1+\kappa^2 \zeta \phi^2
 (1+6 \zeta)}\right)  \frac{ \dot{\phi}^2} {\phi}  =  
      \frac{ \kappa^2 \zeta \phi V(\phi) - ( 1+ \kappa^2 
      \zeta \phi^2)V_{,\,\phi}(\phi)}{1+\kappa^2 \zeta \phi^2 (1+6\zeta)} \,,
\end{eqnarray}
that in the slow-roll approximation 
( $|\dot\phi/\phi| \ll H$ and $|\dot{\phi}^2| \ll V(\phi)$ ) can be written as:
\begin{eqnarray}
\label{E1}
H^2 & \simeq & \frac{\kappa^2}{3(1+\kappa^2\zeta\phi^2)}V(\phi) \,, \\
\label{E2}
3 H \dot{\phi} & \simeq &   
\frac{ \kappa^2 \zeta \phi V(\phi) - ( 1+ \kappa^2 \zeta \phi^2)V_{,\,\phi}(\phi) }
{1+ \kappa^2 \zeta \phi^2 (1+6\zeta)} \,.
\end{eqnarray}
{\it Cosmological perturbations.}
Introducing the following quantities 
\begin{eqnarray}
\hat{Q_S}  =  \left( \frac{ {\rm d} \hat{\phi} / {\rm d} \hat{t} } {\hat{H}} \right)^2 \,,
\hspace{1cm}
\hat{\epsilon}= -\frac{ {\dot {\hat H}} }{\hat{H}^2}\,, 
\hspace{1cm} 
\hat{ \delta}_S  =  \frac{ {\dot {\hat Q}}_S} {2 \hat{H} \hat{Q}_S } \,, \nonumber \\
{\hat {\gamma}_S}=\frac{(1+{\hat \delta}_S)(2+{\hat \delta}_S)+{\hat \epsilon})}
{(1+{\hat \epsilon})^2}\,,
\hspace{1cm} {\hat \nu}_S=\sqrt{{\hat \gamma}_S+1/4}\,,
\end{eqnarray}
the power spectrum of the curvature perturbations in the Einstein frame is given by \cite{Komatsu99,Tsu04}:
\begin{equation}
\label{PS}
{\cal P}_{\cal R}=\frac{1}{ {\hat Q}_S}\left(\frac{{\hat H}}{2 \pi}\right)^2
\left(\frac{1}{{\hat a}{\hat H}|\tau|}\right)^2
\left( \frac{\Gamma({\hat \gamma}_S)}{\Gamma(3/2)}\right)^2
\left(\frac {k |\tau|}{2}\right)^{3-2{\hat \nu}_S}\,.
\end{equation}
where $\tau=-1/[(1+{\hat \epsilon}{\hat a}{\hat H}]$ is the conformal time. \\
The spectral index of the curvature perturbations 
obtained as usual as $n_S \equiv 1+ d\,{\rm ln} {\cal P}_S(k)/ d\,{\rm ln} k$ is:
\begin{equation}
n_S-1=3-2 {\hat \nu}_S=3-\sqrt{4{\hat \gamma}_S+1}\,.
\end{equation}
The power spectrum of the tensor perturbations in the Einstein frame is obtained from Equation (\ref{PS}) making  the following replacements:
\begin{eqnarray}
{\hat Q}_S \rightarrow {\hat Q}_T =1 \,, 
\hspace{0.5cm}
{\hat \gamma}_S \rightarrow {\hat \gamma}_T=\frac{(1+{\hat \delta}_T)(2+{\hat \delta}_T)+{\hat \epsilon})}
{(1+{\hat \epsilon})^2}\,,
\hspace{0.5cm}
\hat{ \delta}_S \rightarrow  {\hat \delta_T}= 
\frac{ {\dot {\hat Q}}_T} {2 \hat{H} \hat{Q}_T }=0 \,.
\end{eqnarray} 
The spectral index of the tensor perturbations $n_T$ and the tensor-to-scalar ratio $R$ are then given by:
\begin{eqnarray}
n_T=3-\sqrt{4{\hat \gamma}_T+1}\, 
\hspace{1cm}
R= 8\frac{{\hat Q}_S}{ {\hat Q}_T } 
\left( \frac{\Gamma ({\hat \gamma}_T)} {\Gamma({\hat \gamma}_S)}\right)^2 \,.
\end{eqnarray}

\subsection{Higgs driven G-inflation}
\label{SS1}

In this section we consider that inflation is driven by the SM Higgs potential
with a kinetic term  modified by a function $G(\phi,X)$ such that the standard part of the Lagrangian given in Equation (\ref{L_SM}) remains much larger than the Galileon 
interaction term, taking the form \cite{Koba10}:
\begin{eqnarray}
\label{L_G}
{\cal L}_{G}(\phi,X)=P(\phi,X)-G(\phi,X)\Box \phi \,,
\hspace{0.3cm}{\rm where}\hspace{0.3cm}
|P(X,\phi)|\simeq V(\phi)\gg |G(X,\phi)\Box \phi|\,.
\end{eqnarray} 
{\it Background equations.} The energy density and pressure of the homogeneous and isotropic background 
field described by the FRW metric are \cite{Kama10,Koba11}:
\begin{eqnarray}
\label{dens}
\rho & = & 2P_{,X}X-P+6G_{,X}H \dot{\phi}X-G_{,\phi}X\,,\\
\label{pres}
p & = & P-(G_{,\phi}+G_{,X}\ddot{\phi})X\,,
\end{eqnarray}
where $H= \dot{a}/a$ is the Hubble parameter. The gravitational field  equations are then given by:
\begin{eqnarray}
\label{grav_eq}
3 \kappa^2 H^2=\rho\,, \hspace{1.cm} -\kappa^2 \left(2H^2+2\dot{H}\right)=p \,,
\end{eqnarray}
and the scalar field equation of motion reads as:
\begin{eqnarray}
\label{field_eq}
 P_{,X}(\ddot{\phi}+3H\dot{\phi}) +2P_{XX}X \ddot{\phi}+2P_{,X\phi}X-P_{,\phi}
-2G_{,X\phi}X(\ddot{\phi}-3H\dot{\phi}) \\ \nonumber
 + 6G_{,X}[(HX)+3H^2X]-2G_{,\phi\phi}X+6G_{,XX}HX\dot{X}=0  \,.
\end{eqnarray}
Since we are interesting in fluctuations generated during inflation, we  assume a background  for which the quantities $|\ddot{\phi}/(H\dot{\phi})|$ and  
$|G_{,\phi} \dot{\phi}/(GH)|$ are much smaller than unity. Under these assumptions, the first slow-roll parameter can be written as \cite{Mizuno10}:
\begin{equation}
\label{eps_G}
\epsilon  \equiv -\frac{\dot{H}}{H^2}=
\frac{XP_{,X}+3G_{,X}H \dot{\phi} X}{\kappa^2 H^2} \ll 1 \,,
\end{equation} 
and the sound speed of scalar perturbations is given by:
\begin{equation}
c^2_s \equiv \frac{\partial_X \, p}{\partial_{X}\,\rho}= 
\frac{\cal F}{\cal G}\,,
\end{equation}
where:
\begin{eqnarray}
{\cal F} & = &  P_{,X}+4 \dot{\phi}H\,G_{,X} \,, \\
{\cal G} &  = &  P_{,X}+2XP_{,XX}+6H \dot{\phi}(G_{,X}+XG_{,XX})\,.
\end{eqnarray}
{\it Cosmological perturbations.} The CMB temperature anisotropy directly relates to the 
statistical properties of the primordial curvature perturbations ${\cal R}$ 
on large scales:  
\begin{equation}
 {\cal R}=-\frac{H}{\dot{\phi}}Q \,,
\end{equation}
where $\phi$ is the background value of the field and $Q$ is the perturbation  
in the unitary gauge, $\phi({\bf x},t)=\phi(t)$, with $\delta \phi=0$. 
The second order action in this gauge at  leading order in the slow-roll approximation  reads as:
\begin{eqnarray}
\label{action2} 
S_2=\frac{1}{2 \pi^2}\int{\rm d}{\tau}{\rm d}^3x\frac{a^3}{2 c^2_s}
(P_{,X}+4 {\dot \phi}HG_{\,X})\left[{\dot Q}^2
 -\frac{c^2_s}{a^2}\partial^i Q \partial_i Q \right]\,,
\end{eqnarray}
where $\tau$ is the conformal time. 
One should note that in the case of G-inflation the unitary gauge 
does not coincide with the comoving gauge \cite{Koba10} 
as in the case of K-inflation  
for which the scalar field Lagrangian is  $L_{\phi}=P(X,\phi)$ \cite{Garriga99}.
Consequently, there are two independent solutions of the perturbation equation resulting from the second order action (\ref{action2}). 
Introducing the following parameters
\begin{eqnarray}
\nu \equiv \frac{G_{,X} {\dot \phi} X}{\kappa^2 H}\,, \hspace{1.cm}
{\tilde \epsilon} \equiv \epsilon + \nu \,,
\end{eqnarray}
the power spectra for $Q$ and ${\cal R}$ are obtained as \cite{Mizuno10}:
\begin{eqnarray}
P_{Q}=\frac{X}{4 \pi^2 M^2_{pl}c_s \tilde{\epsilon}}\,,
\hspace{1.5cm} P_{\cal R}=\frac{1}{8\pi^2M^2_{pl}}\frac{H^2}{c_s \tilde{\epsilon}}\,,
\end{eqnarray}
which are evaluated at the time of sound horizon exit $c_sk=aH$. 
Defining the additional slow-roll parameters: 
\begin{eqnarray}  
\tilde{\eta} \equiv \frac{ \dot{\tilde{\epsilon}}} {\tilde{\epsilon}H}\,,
\hspace{1cm} s\equiv\frac{\dot{c_s}}{c_sH}\,,
\end{eqnarray}
the spectral index of the curvature perturbations is obtained as:
\begin{equation}
n_s-1=\frac{{\rm d}P_{\cal R}(k)}{ {\rm d \, ln}k}=-2\epsilon-\tilde{\eta}-s  \,.
\end{equation}
The power spectrum and the spectral index of tensor perturbations are given as usual in the form:
\begin{eqnarray}
P_{T}=\frac{2H^2}{\pi^2M^2_{pl}}\,, \hspace{1cm} n_{T}=-2\epsilon \,.
\end{eqnarray}
Thus, the tensor-to-scalar ratio $R$ is:
\begin{equation}
R\equiv \frac{P_{T}}{P_{\cal R}}=-8 c_s (n_{T}-2 \nu)\,,
\end{equation}
that highlights the difference between G-inflation model 
and K-inflation model for which $R=-8 c_s n_{T}$.

Hereafter we will consider Higgs potential driven G-inflation corresponding
to the following setup \cite{Kama10}:
\begin{eqnarray}
\label{G_eq}
G(\phi,X)=-g(\phi)X\,, \hspace{1.5cm} g(\phi)=\frac{\phi}{w^4} \,, 
\end{eqnarray}
where $w$ is a mass parameter assumed to be positive. Under the slow-roll conditions, the energy density is dominated by the potential energy and
the standard part of the kinetic term in the Lagrangian given in equation
(\ref{L_G}) takes the form $ |P(\phi,X)|\simeq V(\phi)$. The speed of sound  
corresponding to this setup is $c^2_s \simeq 2/3$ and 
the Friedmann equation and the equation of motion of the scalar field read as:
\begin{eqnarray}
\label{Friedmann}
H^2    \simeq  \frac{\kappa^2}{12} \lambda \phi^2 \,,  \hspace{1.5cm}
{\dot \phi^3}  \simeq  -\frac{2}{3} \frac{\kappa^2 H}{g(\phi)}\epsilon \,.
\end{eqnarray}

\subsection{Higgs driven E-inflation}
In the recently proposed SM Higgs inflation model (E-inflation) 
with non-minimal derivative coupling to gravity \cite{Germani10a}, the tree-level SM Higgs Lagrangian takes the form:
\begin{eqnarray} 
\label{L_C}
{\cal L}_{E}(X,\phi) =-\frac{1}{2}\left(g^{\mu \nu}-w^2G^{\mu \nu}\right)\partial_{\mu} \phi \partial_{\nu}\phi - V(\phi)\,, \hspace{0.5cm} 
G^{\mu \nu}=R^{\mu \nu}-\frac{\bf R}{2}g^{\mu \nu}\,,
\end{eqnarray}
where ${\bf R}$ is the Ricci scalar, $G^{\mu \nu}$ is the Einstein tensor, $w$ is a mass parameter and $V(\phi)$ is the Higgs potential. \\ 
{\it Background equations.} The Friedmann equation and the field equation of motion are given by:
\begin{eqnarray}
\label{eq_back}
H^2=\frac{\kappa^2}{6}\left[\dot{\phi}^2(1+9H^2w^2)+\frac{\lambda}{2}
\phi^4\right] \,, \hspace{0.5cm}
\partial_t\left[a^3\dot{\phi}(1+3H^2w^2)\right]=-a^3\lambda\phi^3 \,. 
\end{eqnarray}
Asking the solutions to obey the following constraints:
\begin{eqnarray} 
\epsilon \equiv-\frac{\dot{H}}{H^2} \ll 1\,, \hspace{0.5cm}
9H^2w^2 \dot{\phi}^2 \ll \frac{\lambda}{2}\phi^4 \,,
\hspace{0.5cm} H \gg \frac{1}{3w}\,, 
\hspace{0.5cm}|\ddot{\phi}| \ll3H|\dot{\phi}| \,,
\end{eqnarray}
Equations (\ref{eq_back}) can be written as:
\begin{eqnarray}
\label{Hub_c}
H^2  \simeq  \frac{k^2}{12} \lambda \phi^4 \,, 
\label{field_c}
\hspace{1.5cm}
\dot{\phi}  \simeq  -\frac{4}{3Hw^2\kappa^2\phi}\,.
\end{eqnarray} 
{\it Cosmological perturbations.} The power spectrum of the curvature  perturbations evaluated at the time of sound horizon exit $c_s k=aH$ is obtained 
as solution of the perturbation equation resulting from the canonically normalized 
second order action \cite{Germani10b}:
\begin{equation}
{P}_{\cal R}=\frac{\kappa^2H^2}{4 \pi^2 }
\frac{1}{12 w^2 \kappa^2 \dot{\phi}^2}\left(1+ \frac{19}{2}w^2\kappa^2\dot{\phi^2}\right)\,, 
\end{equation}
where the sound speed of  the curvature perturbations is 
\begin{equation}
0 < c^2_s=\frac{3-13w^2\kappa^2 {\dot \phi}^2}
{3+18w^2\kappa^2 {\dot \phi}^2}< 1\,.
\end{equation}
Introducing the first two slow-roll parameters:
\begin{eqnarray}
\label{HSR1_c}
\epsilon & \equiv & -\frac{\dot{H}}{H^2}=\frac{3}{2}\kappa^2w^2\dot{\phi}^2 \hspace{0.5cm}\,,\\
\label{HSR2_c}
\eta & \equiv & -\frac{\ddot{\phi}}{H \dot{\phi}}=-\frac{9}{4}\kappa^2w^2\dot{\phi}^2
=-\frac{3}{2}\epsilon \,,
\end{eqnarray} 
to leading order in slow-roll approximation ${P}_{\cal R}$ can then be written as: 
\begin{equation}
\label{As_c}
{P}_{\cal R}=\frac{\kappa^2H^2}{4 \pi^2}\frac{1}{8\epsilon} \,,
\end{equation}
and the spectral index of the curvature perturbations is
\begin{equation}
n_S-1=\frac{{\rm d \,ln} {\cal P}_{\cal R}}{{\rm d \, ln}k}=2 \frac{{\rm d \, ln }H}{{\rm d \, ln}k}- \frac{{\rm d\,ln}\epsilon}{{\rm d \, ln}k} \,,
\end{equation}
and by using Equations (\ref{HSR1_c}) and (\ref{HSR2_c}) can be written as:
\begin{equation}
\label{ns_c}
n_S=1-2\epsilon-2\eta=1-5\epsilon\,.
\end{equation}
The power spectrum of the tensor perturbations is obtained as:
\begin{eqnarray}
\label{At_c}
P_T=\frac{\kappa^2 H^2}{4 \pi^2 \Omega^2c^3_g}\,,
\end{eqnarray}
where $\Omega$ and the sound speed of the gravitational waves $c_g$ are:
\begin{eqnarray}
\Omega^2=1-\frac{1}{2}w^2\kappa^2 \dot{\phi}^2 \,,
\hspace{0.5cm}c^2_g\simeq 1+w^2\kappa^2 \dot{\phi}^2 \,.
\end{eqnarray}
To the leading order in $w^2\kappa^2 \dot{\phi}^2$ we have $\Omega \simeq 1$ and  $c_g \simeq 1$.
The spectral index of the tensor perturbations is 
\begin{equation}
n_T=\frac{{\rm d \, ln}{P}_T}{{\rm d \, ln}k}=-2 \epsilon \,,
\end{equation} 
and the tensor-to-scalar ratio $R$  is given by:
\begin{equation}
\label{R_c}
R= \frac{{P}_T}{{P}_{\cal R}}=12 w^2 \kappa^2 \dot{\phi}^2 =8 \epsilon \,. 
\end{equation}

\section{Quantum corrections to the Higgs potential}
The quantum corrections due to the interaction effects of 
the SM particles with Higgs boson through quantum loops modify Higgs scalar potential from classical expression in both  Jordan and Einstein frames, taking the RG forms: 
\begin{eqnarray}
\label{pot_rg}
V(t)  =  \frac{\lambda(t)} {4} \phi(t)^4\,,
\hspace{0.5cm}
\hat{V}(t) & = & \frac{1}{16 \kappa^4} \,
\frac{\lambda(t)\phi^4(t)}
{ (1+\kappa^2\xi(t)\phi^2(t))^2}\,,
\hspace{0.5cm}
t= {\rm ln} \left(\frac{\phi}{m_{Top}}\right)\,,
\end{eqnarray}
where the scaling variable $t$ normalizes the Higgs field and all the running couplings to the top quark mass scale $m_{Top}=171.3$ GeV \cite{PDG}. We compute the various $t$-dependent running constants by integrating the RG  $\beta$-functions as compiled in the Appendix.
For each case, the $t$-dependent running constants are obtained as:
\begin{eqnarray}
\label{running} 
Y(t)=\int^{t}_{t=0} \frac{\beta_{Y}(t')}{1+\gamma(t')}\,{\rm d}t' \,,
\hspace{1.cm} Y=\{g,g',g_s,y_t,\lambda, \zeta\}\,, 
\end{eqnarray}
where $\gamma$ is the Higgs field  anomalous dimension given by Equation (\ref{gamma}) from Appendix.\\
At $t=0$, which corresponds to the top quark mass scale $m_{Top}$,  
the Higgs quadratic coupling $\lambda(0)$ and  the top Yukawa coupling $y_t(0)$  are determined by the corresponding pole masses and the vacuum expectation value 
${\it v}=246.22$ GeV as: 
\begin{eqnarray}
\label{pol_mass}
\lambda(0)=\frac{m^2_H}{2 {\it v}^2}\left[1+2 \Delta_H (m_H)\right]\,, 
\hspace{1.5cm}y_t(0)=\frac{\sqrt{2} m_T}{\it v}\left[1+\Delta_T(m_T)\right]\,,
\end{eqnarray}
where $\Delta_H(m_H)$ and $\Delta_T(m_T)$ are the corrections to Higgs and top quark mass respectively, 
computed following the scheme from the Appendix of Espinosa et al.(2008). 
The initial values of the gauge coupling constants at $t=0$ are given by:
\begin{eqnarray}
\frac{g^2(0)}{4 \pi}=0.03344\,,
\hspace{1cm} 
\frac{g'^2(0)}{4 \pi}=0.01027 \,,\hspace{1cm}\frac{g^2_s(0)}{4 \pi}=0.1071\,.
\end{eqnarray}
The couplings $g$ and $g'$ are obtained by integrating RG flow equations from their values at $M_Z$, while $g_s$ is calculated numerically \cite{BarSta09}. 
The value of non-minimally coupling constant $\zeta$ at the beginning of inflation is determined such that the calculated value of the amplitude of the curvature density perturbations agrees with the measured value \cite{Lyth}.

\section{RESULTS}
\subsection{The CMB Angular Power Spectra}

We obtain the CMB temperature anisotropy and polarization power spectra
by integrating the coupled Friedmann equation and equation of motion of the scalar field
with respect to the conformal time corresponding to the SM Higgs inflation variants, 
as presented in the previous section. 
We take wavenumbers in the range $5 \times 10^{-6}-5$~Mpc$^{-1}$
needed by CAMB Boltzmann code \cite{camb,Lewis02} to numerically derive the CMB angular power spectra and a Hubble radius crossing  scale $k_*=c_sk=0.002$Mpc$^{-1}$ and  
impose that the electroweak vacuum expectation value ${\it v}=$246.22 GeV is the true minimum of the Higgs potential at any energy scale ($\lambda(t)>$0).
 
The value of the Higgs scalar field at the beginning of inflation $\phi_{\cal N}$  
determines the value of the scaling parameter $t={\rm ln} (\phi_{\cal N}/m_{Top})$ at this time.\\
For the case of $\zeta$-inflation the value of $\phi_{\cal N}$ 
relates to the quantum scale of inflation $\phi_I$ and to the duration of inflation expressed in units of {\it e}-folding number ${\cal N}$ through \cite{BarSta08}:
\begin{eqnarray}
\frac{\varphi^2_*}{\varphi_I} & = & e^x -  1 \,, 
\hspace{1.5cm}
 \varphi_I  =\frac{64 \pi^2 M^2_{pl}}{\xi {\bf A_I}}\,, 
\hspace{1.5cm} 
x \equiv \frac{{\it N}{\bf A_I}}{48 \pi^2}\,, \\ 
{\bf A_I} & = & \frac{3}{8 \lambda}\left(2g^4+(g^2+g'^2)^2-16y_t^4\right)-6\lambda\,,
\end{eqnarray}
where the {\it inflationary anomalous scaling} parameter ${\bf A_I}$ 
involves a special combination of quantum corrected coupling constants \cite{BarKa94,BarSta09}.
\\
The value of the field at the end of G-inflation is obtained 
from Equations (\ref{eps_G}) and (\ref{G_eq}) asking the solution to 
obey the condition $\epsilon=1$ at the end of G-inflation:
\begin{equation}
\label{phi_end_G}
\phi_{end}= 2^{3/4}\lambda^{-1/8} \kappa w^{1/2} \,.
\end{equation}
The number of $e$-folds till the end of G-inflation is then given by:
\begin{equation}
\label{N_G}
{\cal N}= \int^{\phi_{end}}_{\phi_{\cal N}} \frac{H}{{\dot \phi}} {\rm d}\phi=
\frac{1}{16} \frac{\lambda^{1/2}}{\kappa^2 w^2}\phi^{4} -\frac{1}{2} \,,
\end{equation}
from which the value of the field evaluated ${\cal N}$ $e$-folds before 
the end of G-inflation is:
\begin{equation}
\phi_{\cal N}= (16 {\cal N} +8 )^{1/4} \lambda^{-1/8} \kappa w^{1/2}\,.
\end{equation}
\\
In a similar way, the value of the field at the end of E-inflation is obtained  from  Equations (\ref{Hub_c}) and (\ref{HSR1_c}) by asking 
the solution to obey the condition $\epsilon=1$ at the end of E-inflation: 
\begin{equation}
\label{phi_end}
\phi^6_{end} = \frac{32}{ w^2 \kappa^4 \lambda} \,,
\end{equation}
From  the evaluation of the number of e-foldings $\cal N$ during E-inflation:
\begin{equation}
{\cal N}=\int^{\phi_{end}}_{\phi_{\cal N}}\frac{H}{\dot{\phi}}{\rm d}\phi=
\frac{1}{96} \lambda w^2 \kappa^4 \left(\phi^6_{\cal N}-\phi^6_{end}\right) \,,
\end{equation}
we obtain the value of the field at ${\cal N}$ e-folds before the end of E-inflation as:
\begin{equation}
\label{phi0_c}
\phi^6_{{\cal N}} = \frac{32} {w^2 \kappa^4 \lambda}\left(3 {\cal N}+1\right)  \,.
\end{equation}

As the inflationary observables are evaluated at
the epoch of horizon-crossing quantified by the number of {\it e}-foldings $\cal N$
before the end of the inflation at which our present Hubble scale
equalled the Hubble scale during inflation, 
the uncertainties in the determination of $\cal N$
translates into theoretical errors in determination of the inflationary observables \cite{Kinney04,Kinney06}.
Assuming that the ratio of the entropy per comoving interval
today to that after reheating is negligible,
the main uncertainty in the determination of $\cal N$  is
given by the uncertainty in the determination of the reheating
temperature $T_r$ after inflation. \\
The dynamics of the Higgs field during G-inflation show 
that the effect of Galileon-like interaction during reheating is very small and can be 
safely ignored because $g=\phi/w^4$ ($\phi \ll w)$ term is suppressed around the minimum of the potential \cite{Kama10}. \\
In the case of E-inflation the constraint $ w H  \gg 1$ 
suppresses the canonically normalized field $\phi \sim (wH)^{-1}$ that  becomes ineffective during reheating. \\
Recent studies of the reheating after $\zeta$-inflation estimates $T_r$  in the range \cite{BGS09,Bellido09}:
\begin{equation}
\label{Tr}
3.4 \times 10^{13}\,{\rm GeV} < T_{r} < \left(\frac{\lambda}{0.25})\right)^{1/4}
\,1.1 \times 10^{14}\,{\rm GeV} \,,
\end{equation}
that translates into a negligible variation of the number of {\it e}-foldings  with the Higgs mass ($\Delta N \sim 0.1$). The number of {\it e}-foldings at Hubble radius crossing  scale $k_*$ is related to $T_r$ through \cite{Lerner11}:
\begin{equation}
\label{N_Tr}
{\cal N}={\rm ln}\left[  \left(\frac{\rho_R}{\rho_{end}}\right)^{1/3}
\left( \frac{g_0T^3_0}{g_*T^3_r}\right)^{1/3}H \lambda\right]\,,
\end{equation}
where $g_* \simeq 106.5$ for the SM, $g_0 \simeq 2$ and $T_0$ is the present value of the photon temperature. Taking the variation range of $T_r$  given in Equation (\ref{Tr}) and ${\cal N}=$ 59 ${\it e}$-foldings at $k_*$, in the view of WMAP7+SN+BAO normalization at this scale, we obtain $57.8 \le {\cal N} \le 60.2$ for 
the number of ${\it e}$-foldings at which  $k_*$ exits the horizon.\\
Throughout, we will consider ${\cal N}$ varying in the above 
range for all SM Higgs inflation variants. In this way we conservatively included 
the same error $\Delta {\cal N} \simeq 1.17$ to account for possible uncertainties 
in the theoretical estimate of $T_r$.\\
For each wavenumber $k$ our code integrates the $\beta$-functions of the  $t$-dependent running constant couplings in the observational inflationary window imposing that
$k$ grows monotonically to the wavenumber $k_*$,
eliminating at the same time those models violating the condition for inflation 
$\epsilon \equiv -{\dot H}/H^2 \le 1$. 

\subsection{Markov Chain Monte Carlo (MCMC) Analysis}

We use MCMC technique to reconstruct the Higgs field 
potential and to derive constraints on the inflationary observables and the Higgs boson mass 
from the following datasets.\\
The WMAP 7-year data \cite{Komatsu10,Larson} complemented  with
geometric probes  from the Type Ia supernovae (SN) distance-redshift relation and
the baryon acoustic oscillations (BAO). 
The SN distance-redshift relation has been studied in detail in the recent
unified analysis of the published heterogeneous SN data sets -
the Union Compilation08  \cite{Kowalski,Riess09}. 
The BAO in the distribution of galaxies  are extracted from Two Degree Field Galaxy Redshift Survey (2DFGRS)
the Sloan Digital Sky Surveys Data Release 7 \cite{Percival09}.
The CMB, SN and BAO data (WMAP7+SN+BAO) are combined by multiplying the likelihoods.
We use these measurements especially because we are testing models deviating from  the  standard Friedmann expansion. These datasets properly enables us to account for any shift of the CMB
angular diameter distance  and of the expansion rate of the universe. \\
The likelihood probabilities are evaluated  by using
the public packages {\sc CosmoMC} and {\sc CAMB}
\cite{Lewis02,camb} modified to include the 
formalisms for the SM Higgs driven  G-inflation and  E-inflation as 
described in the previous sections. \\
The fiducial model is the $\Lambda$CDM standard cosmological model 
described by a set of parameters receiving uniform priors.
For the case of G-inflation and E-inflation we used the following set of parameters:
$$\left\{ \Omega_bh^2 \,,\,\,\Omega_ch^2\,,\,\, \theta_s\,,\,\, \tau\,,\,\,
m_{Higgs}\,,\,\, w \,,\,\, {\cal N}\right\} \,,$$
where: $\Omega_{b}h^2$ is the 
physical baryon  density, $\Omega_ch^2$ 
is the physical dark matter density,  $\theta_s$
is the ratio of the sound horizon distance 
to the angular diameter distance,  $\tau$ is 
the reionization optical depth,  
$m_{Higgs}$ is the Higgs boson pole mass, $w$ is the mass parameter 
and $\cal N$ is the number of {\it e}-foldings at the Hubble radius crossing.\\
For the case of $\zeta$-inflation we used  the same set 
of input parameters except for the mass parameter $w$. For this case we use as input  the amplitude of the curvature perturbations $A^2_S \sim \lambda /\zeta^2$ that fix the value of the non-minimally coupling constant $\zeta$ at $k_*$ for a given value of the Higgs boson mass. \\
For each inflation model we run 64 Monte Carlo Markov chains, imposing for each case the Gelman \& Rubin convergence criterion \cite{Gelman92}.
\begin{figure}
\begin{center}
\includegraphics[height=12cm,width=12cm]{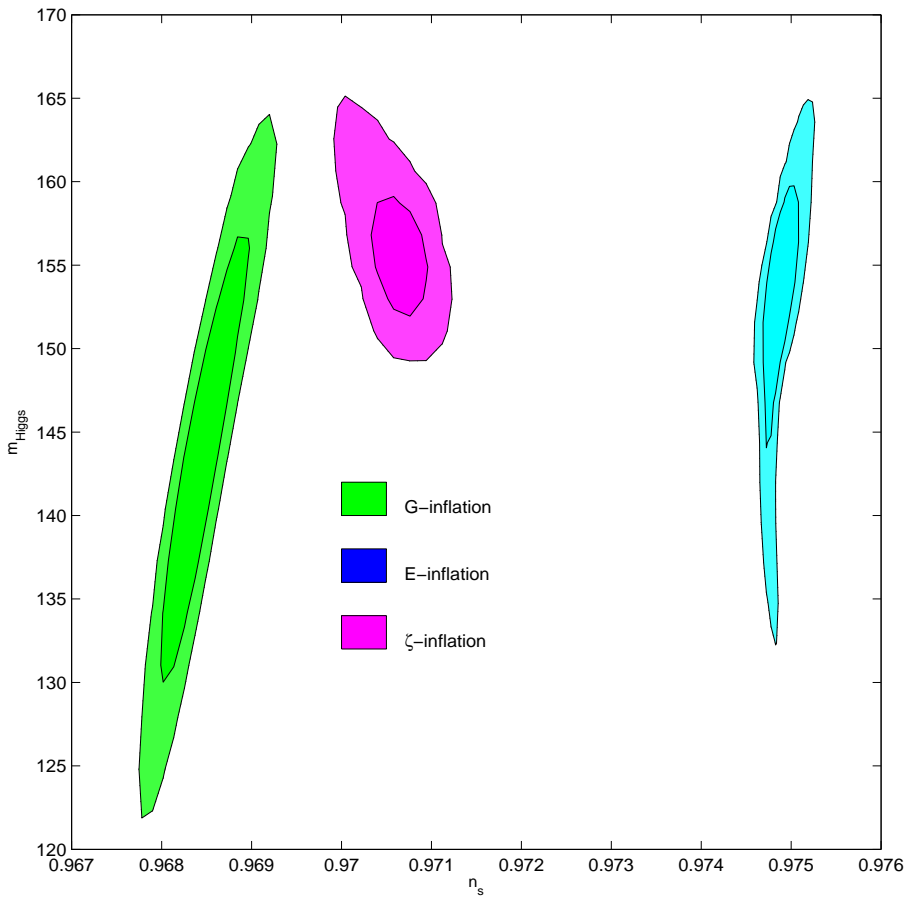}
\end{center}
\caption{Two-dimensional marginalized joint probability distributions
(68\% and 95\% CL) showing the dependence of the spectral index of the curvature 
perturbations $n_S$ on the Higgs inflaton mass $m_{Higgs}$ as obtained from 
the fit  of  the SM Higgs inflation variants  
with $m_{Top}$=171.3 GeV and ${\it v}$=246.22 GeV to the WMAP7+SN+BAO data set.
All parameters are computed at the Hubble crossing scale $k_*$=0.002Mpc$^{-1}$.}
\end{figure}
\begin{figure}
\begin{center}
\includegraphics[height=10cm,width=16cm]{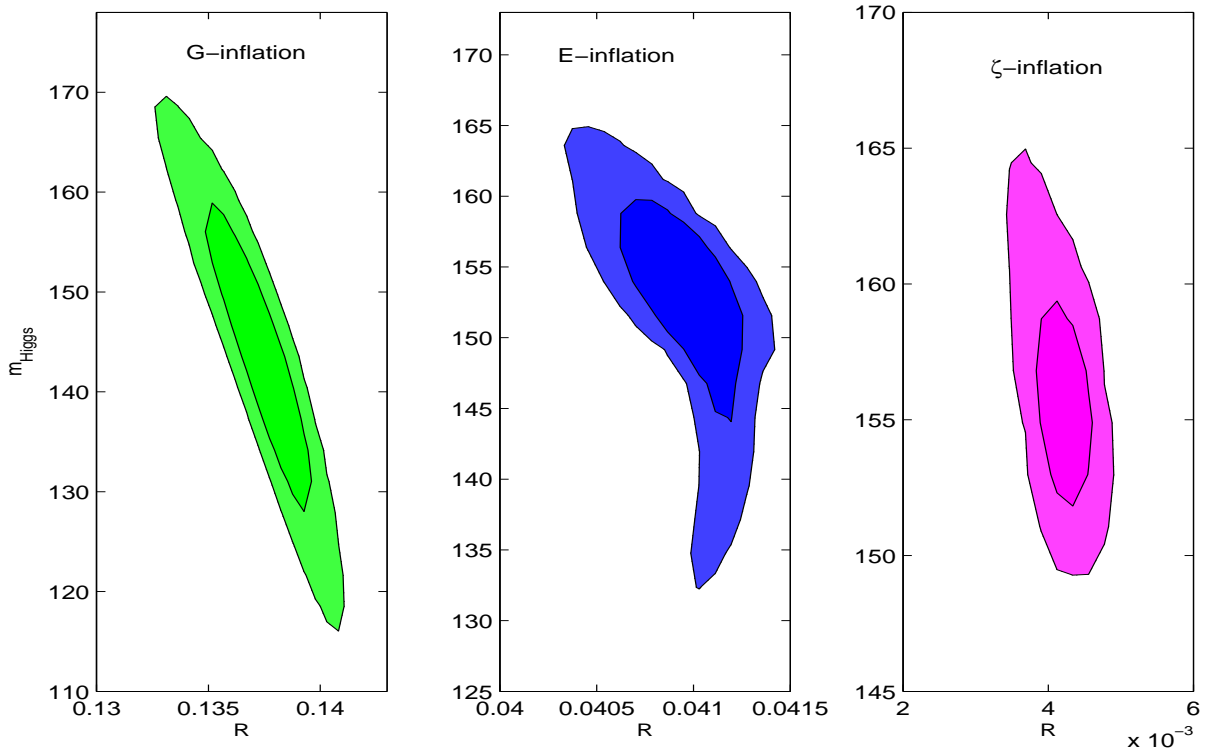}
\end{center}
\caption{Two-dimensional marginalized joint probability distributions
(68\% and 95\% CL) showing the dependence of the tensor-to-scalar ratio $R$ 
on the Higgs inflaton mass $m_{Higgs}$ as obtained from 
the fit  of  the SM Higgs inflation variants  
with $m_{Top}$=171.3 GeV and ${\it v}$=246.22 GeV to the WMAP7+SN+BAO data set.
All parameters are computed at the Hubble crossing scale $k_*$=0.002Mpc$^{-1}$.}
\end{figure}
\begin{figure}
\begin{center}
\includegraphics[height=10cm,width=16cm]{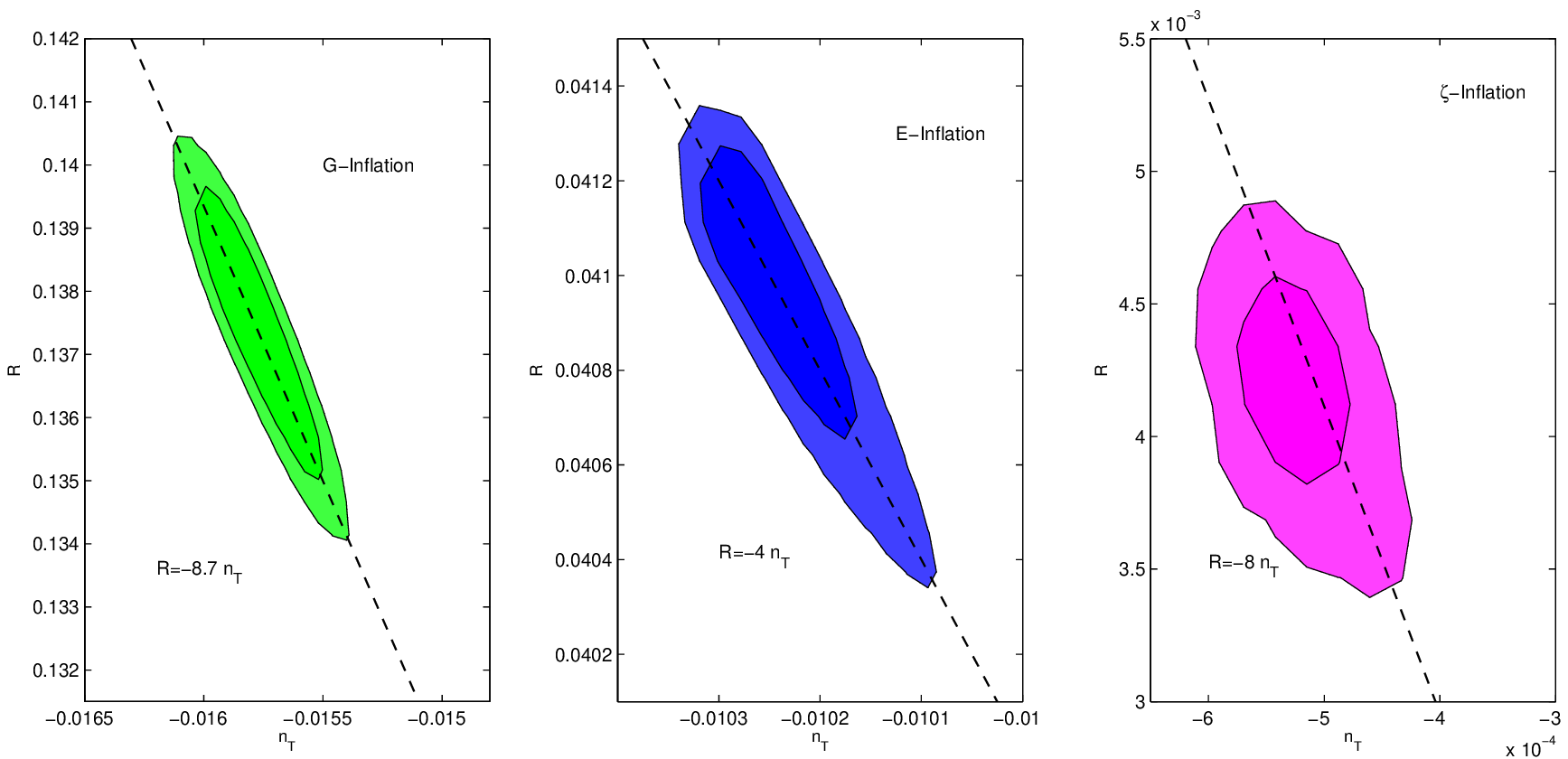}
\end{center}
\caption{Two-dimensional marginalized joint probability distributions (68 \% and 95\% CL) showing 
the degeneracy between the spectral index of tensor perturbations $n_T$ and the tensor-to-scalar 
ratio $R$ obtained from the fit of SM Higgs inflation models with $m_{Top}$=171.3 GeV and ${\it v}$=246.22 GeV 
to the WMAP7+SN+BAO dataset. For each case the dashed 
line shows the theoretical consistency relation. 
All parameters are computed at the Hubble crossing scale $k_*$=0.002Mpc$^{-1}$.}
\end{figure}

Figure~1 and Figure~2 show the dependence of the spectral index of the curvature perturbations and of the tensor-to-scalar ratio respectively on the Higgs boson mass, as obtained from the fit  of  the SM Higgs inflation variants  to the WMAP7+SN+BAO data set.
Table~1 presents the mean values and the lower and upper intervals of the input and derived parameters as obtained from their posterior distributions corresponding to each model. 

We find that the confidence regions of $n_S$ and  $m_{Higgs}$  for  G-inflation and E-inflation models are correlated as $A^2_S \sim \lambda$ increases for heavier Higgs boson.
For $\zeta$-inflation model we find the same confidence regions anticorelated.  
Since for this case we fixed the non-minimal coupling constant $\xi$ such that the amplitude of the curvature perturbations $A^2_S \sim \lambda/\xi^2$ is at the observed value, 
this anticorrelation reflects the fact that in the inflationary domain
the running of $\zeta$ dominates over the running of $\lambda$.
Figure~1 clearly shows that at 2-$\sigma$ level the variation ranges of $n_S$ obtained for SM  Higgs inflation variants do not overlap. 
However, to discriminate among different models the measurement of the Higgs 
boson mass and an improved detection of $n_S$ are required. \\  
The connection between the physics of inflation and the low energy physics expected to be proved by the collider experiments depends on the choice of  UV-completion of each theory.
For $\zeta$-inflation and $E$-inflation this introduces an uncertainty in the determination of Higgs inflaton mass \cite{Bez11,Bez10}. For the case of G-inflation it has recently been argued
that the choice of the Lagrangian in the form $L=G(X,\phi)\box \phi$ leads to second order equations of motion for any choice of $G(X,\phi)$ and therefore to unitarity evolution 
as quantum field theory \cite{Puj}.\\
Therefore, if PLANCK satellite, designed to measure $n_S$ to 2-$\sigma$ accuracy 
of $4 \times 10^{-4}$, will find $n_S$ smaller then 0.97,  while LHC 
will find a Higgs mass significantly smaller then 151 GeV, the $\zeta$-inflation and $E$-inflation models would be ruled out. 
On the other hand, if PLANCK will find $n_S$ 
higher then 0.97 while LHC will find the Higgs mass significantly higher then 151 GeV
the G-inflation model would be ruled out.
One should note that the actual value of $n_S$  measured by the WMAP team at 68\% CL is 
$n_S = 0.968 \pm 0.012$ \cite{Komatsu10}.

Figure~2 presents the equivalent dependences for $R$. The joint confidence regions 
of $R$ and $m_{Higgs}$ are uncorrelated for all SM Higgs inflation variants. These can be attributed to a smaller contribution of the tensor modes to the primordial density perturbations when the Higgs mass is increased. 
We also find a striking difference among the predicted intervals for $R$ by different models. While in the case of $E$-inflation and $\zeta$-inflation the predicted values for $R$ are very low, the G-inflation model predicts $R$ values 
large enough to be detected by the PLANCK satellite. Also for this case we conclude that the measurement by LHC of the Higgs boson mass and the detection by PLANCK of the tensorial modes are required to discriminate among the SM Higgs inflation variants. If  PLANCK will detect a value of  $R \sim 0.1$ while  LHC will find a Higgs mass significantly smaller then 151 GeV, the $\zeta$-inflation and $E$-inflation models would be ruled out. 

Figure~3 presents  the dependence of the tensor-to-scalar ratio on the
spectral index of tensor perturbations (the consistency relation) as obtained from the fit  of  the SM Higgs inflation variants  to the WMAP7+SN+BAO data set. Figure~4 clearly shows that the consistency relations are unique enough to distinguish 
the SM Higgs inflation variants.
\begin{figure}
\begin{center}
\includegraphics[height=10cm,width=16cm]{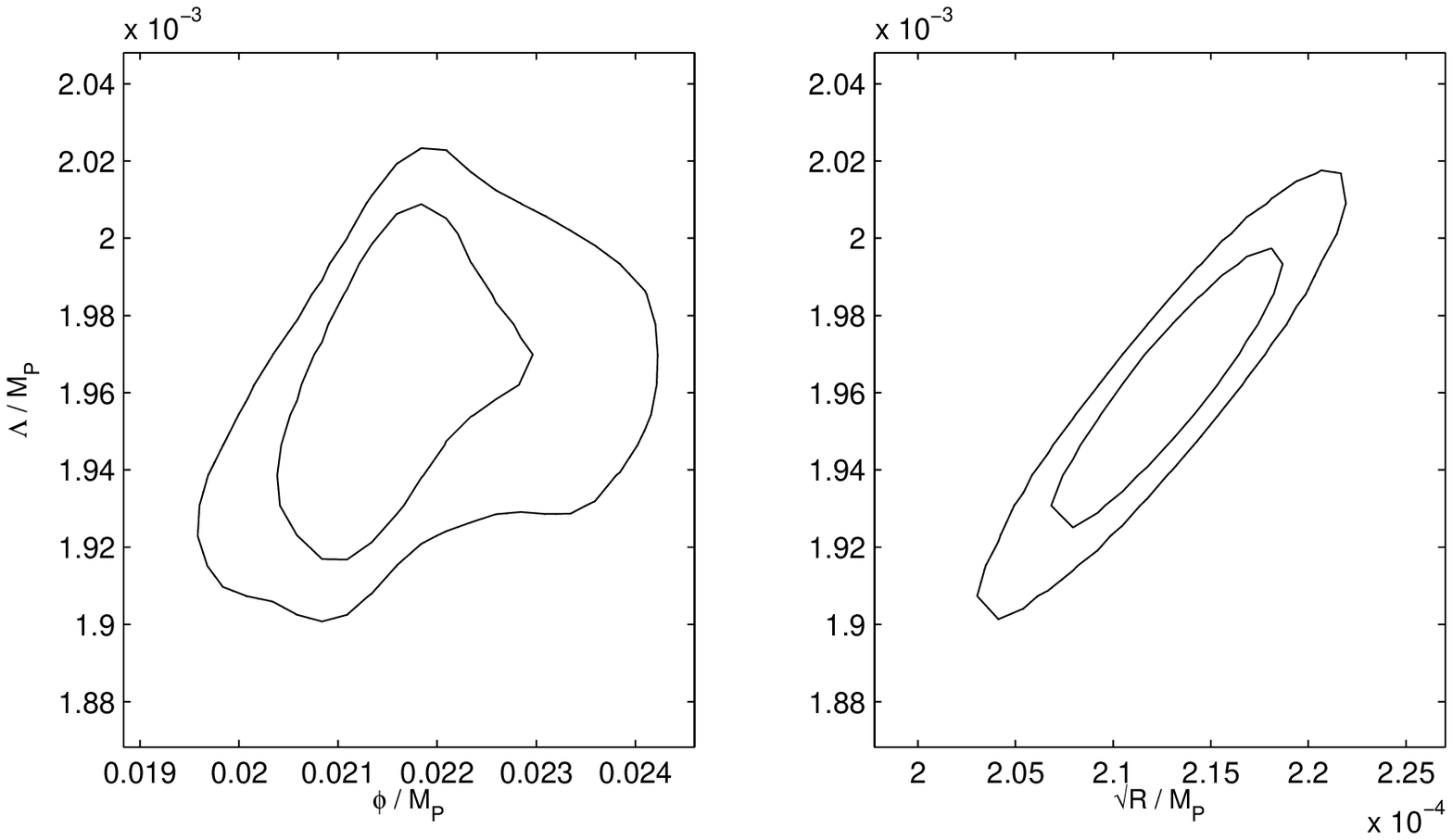}
\end{center}
\caption{Two-dimensional marginalized joint probability distributions
(68\% and 95\% CL) showing the dependence of the scale at which unitarity is violated $\Lambda$ 
on the value of the background field $\phi$ (left) 
and on the comoving curvature perturbations ${\cal R}$ (right) as obtained from 
the fit  of  the E-inflation model  
with $m_{Top}$=171.3 GeV and ${\it v}$=246.22 GeV to the WMAP7+SN+BAO data set.
All parameters are computed at the Hubble crossing scale $k_*$=0.002Mpc$^{-1}$.}
\end{figure}

Finally, we address the question of the violation of perturbative unitarity for E-inflation model. Figure~4 presents the dependence of the scale at which unitarity is violated in this model,
$\Lambda \simeq (2H^2 \kappa)^{1/3}$, on the value of the background field $\phi$ 
and on the comoving curvature perturbations ${\cal R}\simeq 12H^2$, as obtained from 
the MCMC analisys. We find that $\Lambda$ 
is smaller than both $\sqrt{R}$ and $\phi$ during E-inflation, confirming that this model 
suffers from unitarity problems \cite{Atkins_b}, contrary to the original claims \cite{Germani10b}.
\begin{table}
\caption{Mean values and the lower  and upper intervals (at 68\% CL) 
of parameters obtained from the fit of the SM Higgs inflation variants  
with $m_{Top}$=171.3 GeV and ${\it v}$=246.22 GeV to the WMAP7+SN+BAO data set.
All parameters are computed at the Hubble radius crossing $k_*$=0.002 Mpc$^{-1}$.}
\begin{center}
\begin{tabular}{lccc}
\hline \hline \\
Model &  $\zeta$-inflation   &   G-inflation & E-Inflation  \\
Parameter &                  &           &        \\
\hline \\
$100\Omega_bh^2$& 2.257$_{2.206}^{2.308}$ & 2.272$_{2.247}^{2.296}$&      2.254$_{2.294}^{2.315}$ \\
$\Omega_ch^2   $&0.114$_{0.111}^{0.117}$  & 0.117$_{0.113}^{0.123}$&      0.118$_{0.114}^{0.121}$  \\ 
$\tau$          & 0.086$_{0.073}^{0.093}$ & 0.094$_{0.083}^{0.109} $&         0.091$_{0.081}^{0.109}$   \\
$\theta_s$      & 1.037$_{1.035}^{1.039}$ & 1.039$_{1.038}^{1.041}$&        1.040$_{1.039}^{1.044}$ \\
$m_{Higgs}$(GeV)&  155.37$_{151.52}^{159.22}$& 143.83$_{134.87}^{152.98}$     &151.578$_{147.76}^{156.46}$  \\
$\zeta \times 10^{-4}$& \,\,3.147$_{2.638}^{3.656}$& ...& ... \\
$w/M_P$& ... &   6.58$_{6.22}^{6.93}$ $\times 10^{-6} $& 4.369$_{4.194}^{4.522} \times 10^{8}$ \\
${\cal N}$ & $58.88^{59.64}_{58.26}$& $59.01_{58.13}^{59.79}$& $58.87_{57.86}^{59.68}$\\
\hline
${\rm ln}[ 10^{10}A^2_S ]$& 3.161$_{3.129}^{3.193}$&3.177$_{3.154}^{3.203}$&  3.151$_{3.124}^{3.188}$  \\
$n_S$ &  \,\,0.972$_{0.9716}^{0.9722}$& 0.968$_{0.9597}^{0.9690}$          &  0.975$_{0.9741}^{0.9795}$  \\
$n_T \times 10^2$& -0.0442$_{-0.0463}^{-0.0421}$ & -1.575$_{-1.592}^{-1.560}$     &-1.023$_{-1.028}^{-1.018}$  \\ 
R &     \,\,\,\,0.0036$_{0.0027}^{0.0045}$&0.137$_{0.134}^{0.140}$              &0.041$_{0.0387}^{0.0421}$\\
\hline \hline
\end{tabular}
\end{center}
\end{table}

\section{CONCLUSIONS}

Although the SM Higgs inflaton non-minimally coupled to gravity can account for inflation ($\zeta$-inflation), other variants of  Higgs inflation within the SM  
based on scalar field theories with non-canonical kinetic term such as
Galileon-like kinetic term (G-inflation) and kinetic term with non-minimal derivative coupling to the Einstein tensor (E-inflation) have been proposed. 
In this paper we have analyzed the possibility to distinguish 
$\zeta$-inflation, G-inflation and E-inflation models through their predictions for  $n_S$ and $R$ as a function  of $m_{Higgs}$ as well as through their predictions for 
the consistency relations.\\
In order to ensure consistent results, we have studied the SM Higgs inflation variants
considering the full dynamics of the background and perturbations of the Higgs field during inflation and compute the RG improved Higgs potential by including the two-loop $\beta$-functions for the SU(2) $\times$ SU(1) gauge couplings, the SU(3) strong coupling, the top Yukawa coupling, the Higgs quadratic coupling and the Higgs field anomalous dimension. 
Assuming that all SM Higgs inflation variants are consistent theories, we use the MCMC technique to derive constraints on the inflationary parameters and the Higgs boson mass 
from their fit to WMAP7+SN+BAO data set.  We show that the SM Higgs inflation variants 
lead to significant constraints on $n_s$, $R$ and $m_{Higgs}$.\\
From the analysis of the confidence regions in $n_S$-$m_{Higgs}$ 
plane we conclude that in order to discriminate the SM Higgs inflation 
variants by using these parameters the measurement by LHC of $m_{Higgs}$ 
 and an improved detection of $n_S$ are required.  
If PLANCK satellite, designed to measure $n_S$ to an accuracy 
of $4 \times 10^{-4}$, will find $n_S$ smaller then 0.97,  while LHC will find a Higgs mass significantly smaller then 151 GeV, the $\zeta$-inflation and $E$-inflation models would be ruled out. \\ 
We also find a striking difference for the predicted values for $R$ by different models. While for $E$-inflation and $\zeta$-inflation the predicted values for $R$ are low, the G-inflation model predicts $R$ values enough large to be detected by the PLANCK satellite. We conclude that if 
PLANCK will detect a value of  $R \sim 0.1$ while  LHC will find a Higgs mass significantly smaller then 151 GeV, the $\zeta$-inflation and $E$-inflation models would be ruled out.  \\
We also show that the consistency relations are distinct enough to differentiate 
the SM Higgs inflation variants.
 
\begin{acknowledgments}
The author would like to thank to Rose Lerner and Dmitry Gorbunov for helpful discussions.\\
This work  was partially supported by CNCSIS Contract 539/2009 and by 
ESA/PECS Contract C98051.
\end{acknowledgments}

\section{Appendix}
In this appendix we collect the SM renormalization group $\beta$-functions
at renormalization energy scale 
$t={\rm ln}(\phi/m_{Top})$ beyond the top quark mass $m_{TOP}$. \\
All $\beta$-functions and the Higgs field  anomalous dimension $\gamma$ include the Higgs field suppression factor $s(t)$ that for the case of $\zeta$-inflation is given by \cite{deSimone09,Lerner11}:
\begin{equation}
s(t)=\frac{1+\zeta(t)\phi^2(t)}{1+(6\zeta(t)+1)\zeta(t)\phi^2(t)}\,.
\end{equation}
For the case of G-inflation and E-inflation we take $s(t)=1$.

The two-loop $\beta$-functions for gauge couplings $g_i=\{g',g,g_s\}$ are \cite{Espinosa08}:
\begin{eqnarray}
\beta_{g_i} & = & \kappa g_i^3 b_i+\kappa^2 g_i^3\left[\sum_{j=1}^3 B_{ij}g_j^2-s(t)d_i^t y_t^2\right],
\end{eqnarray}
where $\kappa=1/16\pi^2$ and
\begin{eqnarray}
\label{beta_g}
b=((40+s(t))/6,-(20-s(t))/6,-7),\quad 
B & = & \left(
\begin{array}{ccc}
199/18 & 9/2 & 44/3 \\
3/2 & 35/6 & 12 \\
11/6 & 9/2 & -26
\end{array}\right), \nonumber \\ 
d^t=(17/6,3/2,2). 
\end{eqnarray}

For the top Yukawa coupling $y_t$, the two-loop  $\beta$-function  is given by \cite{deSimone09}:
\begin{eqnarray}
&\beta_{y_t}&= 
 \kappa \, y_t \left[-\frac{9}{4} g^2-\frac{17
   }{12}g'^2-8 g^2_s+\frac{9}{2} s(t) y_t^2\right]
+\kappa^2 y_t
   \Bigg{[}-\frac{23}{4} g^4-\frac{3}{4} g^2 g'^2+\frac{1187 }{216}g'^4 + 9 g^2 g_s^2 \nonumber \\
 &+&\frac{19}{9} g'^2 g_s^2-108 g_s^4+
 \left(\frac{225}{16}g^2+\frac{131 }{16}g'^2+36 g_s^2\right) s(t) y_t^2 + 6 \left(-2 s^2(t) y_t^4-2
   s^3(t) y_t^2 \lambda +s^2(t) \lambda   ^2\right)\Bigg{]}\,. \nonumber \\
   \label{beta_yt}
\end{eqnarray}

The two-loop $\beta$-function for the Higgs coupling $\lambda$ is \cite{deSimone09,Lerner11}:
\begin{eqnarray}
\beta_\lambda & = &
 \kappa \left[(18s^2(t)+6) \lambda ^2-6 y_t^4+\frac{3}{8} \left(2 g^4+\left(g^2+g'^2\right)^2\right)+\left(-9 g^2-3
   g'^2+12 y_t^2\right) \lambda \right]\nonumber\\
   &+& \kappa^2
  \Bigg{[}\frac{1}{48} \left(915 g^6-289 g^4 g'^2-559 g^2 g'^4-379 g'^6\right)+30
   s(t) y_t^6-y_t^4 \left(\frac{8 g'^2}{3}+32 g_s^2+3 s(t) \lambda
   \right)\nonumber\\
   &+& \lambda  \left(-\frac{73}{8} g^4+\frac{39}{4} g^2 g'^2+\frac{629
   }{24}s(t) g'^4+108 s^2(t) g^2  \lambda +36s^2(t) g'^2 \lambda -312
   s^4(t) \lambda ^2\right)\nonumber\\
   &+& y_t^2 \left(-\frac{9}{4} g^4+\frac{21}{2} g^2
   g'^2-\frac{19}{4}g'^4+ \lambda  \left(\frac{45}{2}g^2+\frac{85
   }{6}g'^2+80 g_s^2-144 s^2(t) \lambda \right)\right)\Bigg{]}. 
 \end{eqnarray}

The  $\beta$-function for non-minimal coupling 
$\zeta$ is given by \cite{Lerner11}:
\begin{eqnarray}
\beta_{\zeta}=\kappa \left( \zeta+\frac{1}{6}\right)\left(6(1+s(t))\lambda + 6y^2_t -
\frac{3}{2}g'^{2}- \frac{9}{2} g^2 \right) \,.
\end{eqnarray}

Finally, the two-loop Higgs field  anomalous dimension $\gamma$ is given by \cite{deSimone09}:
\begin{eqnarray}
\label{gamma}
\gamma & = &   -\kappa  \left[\frac{9 g^2}{4}+\frac{3 g'^2}{4}-3
   y_t^2\right] 
    -   \kappa^2 \left[ \frac{271
   }{32}g^4- \frac{9}{16} g^2 g'^2 
   -\frac{431}{96} s(t) g'^4   \right]  \nonumber \\ 
  &  + & \kappa^2 \left[-\left(\frac{45}{8}g^2+\frac{85} {24}g'^2+20 g_s^2 \right) y_t^2 
    +  \frac{27}{4} s(t) y_t^4 -6 s^3(t) \lambda ^2\right] \,,
\end{eqnarray}


\begin{thebibliography}{99}

\bibitem{BezSha08} F. Bezrukov, \& M. Shaposhnikov, "The Standard Model Higgs boson as the inflaton" Phys. Lett. B {\bf 659}, 703 (2008), [arXiv:0710.3755 [hep-th]] 

\bibitem{BarSta08} A. O. Barvinsky, A. Yu.  Kamenshchik, 
A. A. Starobinsky, "Inflation scenario via the Standard Model Higgs boson and LHC",   JCAP {\bf 11}, 021 (2008) [arXiv:0809.2104 [hep-ph]]

\bibitem{BerSha09} F. L. Bezrukov, A. Magnin,  M. Shaposhnikov, 
" Standard Model Higgs boson mass from inflation", Phys.  Lett. B {\bf 675}, 88 (2009) [arXiv:0812.4950 [hep-ph]]

\bibitem{deSimone09} de Simone, A., Hertzberg, M. P.,  Wilczek, F. 2009,  
"Running inflation in the Standard Model", Phys. Lett. B {\bf 678}, 1 
[arXiv:0812.4946 [hep-ph]]

\bibitem{BGS09} F. Bezrukov, D. Gorbunov,  M. Shaposhnikov, "On initial conditions for the hot big bang". JCAP {\bf 06}, 029 (2009) [arXiv:0812.3622 [hep-ph]]

\bibitem{BarKa94} A. O. Barvinsky \& A. Yu. Kamenshchik, 
"Quantum scale of inflation and particle physics of the early universe",
Phys. Lett. B {\bf 332},  270 (1994) [arXiv:gr-qc/9404062]

\bibitem{Espinosa08} J. R. Espinosa, G. F. Giudice, A. Riotto, A. 
"Cosmological implications of the Higgs mass measurement",
JCAP {\bf 05}, 002 (2008) [arXiv:0710.2484]

\bibitem{Barbon09} J. L. F.  Barb\'{o}n \& J. R. Espinosa, 
"On the Naturalness of Higgs inflation", Phys. rev. D {\bf 79},
 081302  (2009) [arXiv:0903.0355 [hep-ph]]

\bibitem{Burgess09} C. P. Burgess, H. M. Lee, M. Trott, 
"Power-counting and the validity of the classical approximation during inflation",
JHEP {\bf 09}, 103 (2009) [arXiv:0902.4465 [hep-ph]]

\bibitem{Lerner10a} R. N. Lerner \&  J. McDonald, J. 2010,
"Higgs inflation and naturalness",
JCAP {\bf 04}, 015  (20010) [arXiv:0912.5463 [hep-ph]]   !naturalness

\bibitem{Lerner10b} R. N. Lerner \&  J. McDonald, J.,
"A unitarity-conserving Higgs inflation model",
Phys. Rev. D {\bf 82}, 103525 (2010) [arXiv:1005.2978 [hep-ph]] 

\bibitem{Burgess10} C. P. Burgess, H. M. Lee, M. Trott, 
"On Higgs inflation and naturalness",
JHEP {\bf 07}, 007 (2010) [arXiv:1002.2730 [hep-ph]]

\bibitem{Hertzberg10} M. P. Hertzberg, 
"On inflation with non-minimal coupling",
JHEP {\bf 11}, 023 92010) [arXiv:1002.2995 [hep-ph]]

\bibitem{Bez11} F. Bezrukov, A. Magnin, M. Shaposhnikov, S. Sibiryakov, 
"Higgs inflation: consistency and generalisations",
JHEP {\bf 01}, 016 (2011) [arXiv:1008.5157 [hep-ph]]

\bibitem{Ferrara11} S, Ferrara, R. Kallosh, A. Linde, A. Marrani, A. van Proeyen, 
"Superconformal symmetry, NMSSM, and inflation", 
Phys. Rev. D. {\bf 83} 025008 (2011), [arXiv:1008.2942 [hep-th]]

\bibitem{Atkins_a} M. Atkins \&  X. Calmet, 
"On the unitarity of linearized General Relativity coupled to matter",
Phys. Lett. B {\bf 697}, 37 (2011) [arXiv:1002.0003 [hep-th]]

\bibitem{Picon} C. Armendáriz-Picon, T. Damour, V. Mukhanov, 
"k-Inflation", Phys. Lett. B {\bf 209} (1999) [arXiv:hep-th/9904075]

\bibitem{Ali} M. Alishahiha, E. Silverstein, D. Tong,
"DBI in the sky: Non-Gaussianity from inflation with a speed limit", 
Phys. rev. D {\bf 70}, 123505 (2004) [arXiv:hep-th/0404084]

\bibitem{Easson} D.A. Easson, S. Mukohyama \& B.A. Powell,
"Observational signatures of gravitational couplings in DBI inflation",
Phys. Rev. D {\bf 81}, 023512 (2010)  [arXiv:0910.1353 [astro-ph.CO]]

\bibitem{Amendola} L. Amendola, 
"Cosmology with nonminimal derivative couplings",
Phys. Lett. B {\bf 301} 175 (1993) [arXiv:gr-qc/9302010]

\bibitem{Capo_a} S. Capozziello, G. Lambiase, 
"Nonminimal Derivative Coupling and the Recovering of Cosmological Constant",
Gen. Rel. Grav. {\bf 31}, 1005 (1999) [arXiv:gr-qc/9302010]

\bibitem{Capo_b} S. Capozziello, G. Lambiase, H.-J.Schmidt, 
"Nonminimal Derivative Couplings and Inflation in Generalized Theories of Gravity",
Annalen Phys. {\bf 9}, 39 (2000) [arXiv:gr-qc/9302010]

\bibitem{Granda_a} L. N. Granda,
"Non-minimal Kinetic coupling to gravity and accelerated expansion", 
JCAP {\bf 07}, 006 (2010) [arXiv:0911.3702 [hep-th]]

\bibitem{Granda_b} L. N. Granda, W. Cardona,
"General non-minimal kinetic coupling to gravity", 
JCAP {\bf 07}, 021 (2010) [arXiv:1005.2716 [hep-th]]

\bibitem{Germani10a} C. Germani \&  A. Kehagias
"New Model of Inflation with Nonminimal Derivative Coupling of Standard Model Higgs Boson to Gravity", 
Phys. Rev. Lett. {\bf 105}, 011302 (2010) [arXiv:1003.2635 [hep-ph]]

\bibitem {Germani10b} C. Germani \& A. Kehagias, A. 
"Cosmological perturbations in the new Higgs inflation", JCAP {\bf 05}, 019 (2010) [arXiv:1003.4285 [astro-ph.CO]]

\bibitem{Atkins_b} M. Atkins \&  X. Calmet,
 "Remarks on Higgs inflation", 
Phys. Lett. B {\bf 697}, 37 (2011) [arXiv:1011.4179 [hep-ph]]

\bibitem{Nicolis09} A. Nicolis, R. Rattazzi \&  E. Trincherini,
"Galileon as a local modification of gravity",
Phys. Rev. D {\bf 79}, 064036 (2009) [arXiv:0811.2197 [hep-th]]

\bibitem{Deffayet09a} C. Deffayet, G. Esposito-Farèse \&  A. Vikman, 
"Covariant Galileon", Phys. Rev. D, {\bf 79} 084003 (2009)
[arXiv:0901.1314 [hep-th]]

\bibitem{Deffayet09b} C. Deffayet, S. Deser \&  G. Esposito-Farèse, 
"Generalized Galileons: All scalar models whose curved background extensions maintain second-order field equations and stress tensors",
Phys. Rev. D, {\bf 80}, 064015 (2009) [arXiv:0906.1967 [hep-th]]

\bibitem{Koba10} T. Kobayashi, T., M. Yamaguchi \& J. Yokoyama, 
"Inflation Driven by the Galileon Field", Phys. Rev. Lett. {\bf 105}, 231302 (2010) [arXiv:1008.0603 [hep-th]]

\bibitem{Kama10} K. Kamada, T. Kobayashi, M. Yamaguchi, J. Yokoyama, 
"Higgs G inflation", Phys. Rev. D {\bf 83}, 083515 
 (2010)  [arXiv:1012.4238 [astro-ph.CO]]

\bibitem{Creminelli10} P. Creminelli, A. Nicolis \&  E. Trincherini
"Galilean genesis: an alternative to inflation", JCAP {\bf 11}, 021 (2010)
 [arXiv:1007.0027 [hep-th]]

\bibitem{Mizuno10} S. Mizuno \& K.  Koyama, 
"Primordial non-Gaussianity from the DBI Galileons", 
Phys. Rev. D {\bf 82} 103518 (2010) [arXiv:1009.0677 [hep-th]]

\bibitem{Burrage11} C. Burrage, C. de Rham, D. Seery, A. Tolley,
"Galileon inflation", JCAP {\bf 01}, 014 (2011) [arXiv:1009.2497 [hep-th]]

\bibitem{Felice11a} A. De Felice \& S. Tsujikawa, 
"Primordial non-gaussianities in general modified gravitational models of inflation",
JCAP {\bf 04}, 029 (2011) [arXiv:1103.1172 [astro-ph.CO]]

\bibitem{Koba11} T. Kobayashi, M. Yamaguchi,  \& J. Yokoyama, 
"Primordial non-Gaussianity from G inflation", Phys. rev. D {\bf 83}, 103524 (2011) [arXiv:1008.0603 [hep-th]]

\bibitem{Lerner11} R. N. Lerner, J. McDonald, 
"Distinguishing Higgs inflation and its variants",
Phys. Rev. D {\bf 83}, 123522 (2011) [arXiv:1104.2468 [hep-ph]]

\bibitem{Popa10} L.A. Popa \& A. Caramete,  
"Cosmological Constraints on the Higgs Boson Mass",
 ApJ {\bf 723}, 803 (2010) [arXiv:1009.1293 [astro-ph.CO]]

\bibitem{Felice11b} A. De Felice, S. Tsujikawa, J. Elliston, R. Tavakol, 
"Chaotic inflation in modified gravitational theories" (2011) [arXiv:1105.4685 [astro-ph.CO]]] 

\bibitem{Futamase89} T. Futamase,   \& K. Maeda,
"Chaotic inflationary scenario of the Universe with a nonminimally coupled ``inflaton'' field",
Phys. Rev. D {\bf 39}, 399 (1989)

\bibitem{Komatsu98} E. Komatsu, \&  T. Futamase, 
"Constraints on the chaotic inflationary scenario with a nonminimally coupled ``inflaton'' field from the cosmic microwave background radiation anisotropy",
Phys. Rev. D {\bf 58}, 023004 (1998) [arXiv:astro-ph/9711340].

\bibitem{Komatsu99} E. Komatsu \&  T. Futamase, 
"Complete constraints on a nonminimally coupled chaotic inflationary scenario from the cosmic microwave background",
Phys. Rev. D {\bf 59}, 064029 (1999) [arXiv:astro-ph/9901127]

\bibitem{Tsu04} S. Tsujikawa \&  B. Gumjudpai,
"Density perturbations in generalized Einstein scenarios and constraints on nonminimal couplings from the cosmic microwave background",
Phys. Rev. D {\bf 69}, 123523 (2004) [arXiv:astro-ph/0402185]

\bibitem{Komatsu10} E. Komatsu, et al. (WMAP Collaboration),
"Seven-Year Wilkinson Microwave Anisotropy Probe (WMAP) Observations:  Cosmological Interpretation", ApJS {\bf 192}, 18 (2011) [arXiv:1001.4538 [astro-ph.CO]]

\bibitem{Larson} D. Larson et al. (WMAP Collaboration)
"Seven-Year Wilkinson Microwave Anisotropy Probe (WMAP) Observations:  Power Spectra and WMAP-Derived Parameters",
ApJS {\bf 192}, 16 (2011) [arXiv:1001.4635 [astro-ph.CO]]

\bibitem{Garriga99} J. Garriga \& V. F. Mukhanov, 
"Perturbations in k-inflation",
Phys. Lett. B {\bf 458}, 219 (1999) [arXiv:hep-th/9904176]

\bibitem{PDG} K. Nakamura et al.,
 "The Review of Particle Physics"
J. Phys. G {\bf 37}, 075021 (2010) 

\bibitem{BarSta09} A. O. Barvinsky, A. Yu. Kamenshchik, C. Kiefer, A. A.  Starobinsky, 
C. F. Steinwach
"Asymptotic freedom in inflationary cosmology with a non-minimally 
coupled Higgs fields",
JCAP {\bf 12}, 003 (2009) [arXiv:0904.1698 [hep-ph]]

\bibitem{Lyth}	D. H. Lyth \& A. A. Riotto, 
"Particle physics models of inflation and the cosmological density perturbation",
Phys. Rep. {\bf 314}, 1 (1999) [arXiv:hep-ph/9807278]

\bibitem{camb} A. Lewis, A. Challinor \& A. Lasenby,
"Efficient Computation of Cosmic Microwave Background Anisotropies in Closed Friedmann-Robertson-Walker Models",
ApJ {\bf 538}, 473 (2000)  \footnote{http://camb.info} [arXiv:astro-ph/9911177]

\bibitem{Lewis02} A. Lewis \& S. Briddle S.,
Phys. Rev. D {\bf 66}, 103511 (2002) \footnote{http://cosmologist.info/cosmomc/}
[arXiv:astro-ph/0205436]

\bibitem{Kinney04} W. H. Kinney, E. W. Kolb, A. Melchiorri,
A. Riotto,  
"Inflationary physics from the Wilkinson Microwave Anisotropy Probe",
Phys. Rev D {\bf 69}, 103516 (2004) [arXiv:hep-ph/0305130]

\bibitem{Kinney06} W. H. Kinney \& A. Riotto,
"Theoretical uncertainties in inflationary predictions",
JCAP {\bf 03} 011 (2006) [arXiv:astro-ph/0511127]

\bibitem{Bellido09} J. Garcia-Bellido, D. G. Figueroa \& J. Rubio,  
"Preheating in the standard model with the Higgs inflaton coupled to gravity",
Phys. Rev. D  {\bf 79}, 063531 (2009) [arXiv:0812.4624 [hep-ph]]

\bibitem{Kowalski} M. Kowalski et al. (Supernova Cosmology Project),
"Improved Cosmological Constraints from New, Old, and Combined Supernova Data Sets",
ApJ {\bf 686}, 749 (2008) [arXiv:0804.4142 [astro-ph]]

\bibitem{Riess09} Riess et al.,
"A Redetermination of the Hubble Constant with the Hubble Space Telescope from a Differential Distance Ladder",
ApJ {\bf 699}, 539 (2009) [arXiv:0905.0695 [astro-ph.CO]]

\bibitem{Percival09} W. J. Percival, et al., 
"Baryon acoustic oscillations in the Sloan Digital Sky Survey Data Release 7 galaxy sample",
MNRAS {\bf 401}, 2148 (2010)  [arXiv:0907.1660 [astro-ph.CO]]

\bibitem{Gelman92}A. Gelman \& D. Rubin, 
Statistical Science  {\bf 7}, 457 (1992)

\bibitem{Bez10} F. Bezrukov, D. Gorbunov, 
"Light inflaton hunter's guide", JHEP {\bf 05}, 010 (2010)
[arXiv:0912.0390 [hep-ph]]

\bibitem{Puj}C. Deffayet, O. Pujolas, I. Sawicki, A. Alexander,
"Imperfect dark energy from kinetic gravity braiding",
JCAP {\bf 10}, 026 (2010) [arXiv:1008.0048 [hep-th]]

\end{thebibliography}
\end{document}